\documentclass[10pt,aps,prb,preprint,twocolumn]{revtex4}
\usepackage{amsmath,amssymb}
\usepackage{graphicx}
\usepackage{pslatex}

\begin{document}

\title{Hypergeometric-Gaussian Modes}

\author{Ebrahim Karimi,$^{1}$ Gianluigi Zito,$^{1}$ Bruno Piccirillo,$^{1}$ Lorenzo
Marrucci,$^{2}$ and Enrico Santamato$^{1,*}$}

\affiliation{$^{1}$Dipartimento di Scienze Fisiche Universit\`{a}
degli
Studi di Napoli ``Federico II'', Complesso di Monte S. Angelo, \\80126 via Cintia Napoli, Italy\\
$^{2}$Consiglio Nazionale delle Ricerche-INFM Coherentia, Napoli,
Italy\\$^*$Corresponding author: enrico.santamato@na.infn.it}

\begin{abstract}
We studied a novel family of paraxial laser beams forming an
overcomplete yet nonorthogonal set of modes. These modes have a
singular phase profile and are eigenfunctions of the photon orbital
angular momentum. The intensity profile is characterized by a single
brilliant ring with the singularity at its center, where the field
amplitude vanishes. The complex amplitude is proportional to the
degenerate (confluent) hypergeometric function, and therefore we
term such beams hypergeometric gaussian (HyGG) modes. Unlike the
recently introduced hypergeometric modes~(Opt. Lett. {\textbf 32},
742 (2007)), the HyGG modes carry a finite power and have been
generated in this work with a liquid-crystal spatial light
modulator. We briefly consider some sub-families of the HyGG modes
as the modified Bessel Gaussian modes, the modified exponential
Gaussian modes and the modified Laguerre-Gaussian modes.\newline
{ocis: 050.1960, 230.6120}
\end{abstract}
\maketitle \noindent In the recent years there has been an
increasing interest in laser beams especially tailored to
experiments. In particular, experimentalists are looking for laser
beams which are either nondiffracting or have a definite value of
the photon orbital angular momentum (OAM) along the propagation
direction. Such special laser beams have found useful applications
in optical trapping, image processing, optical tweezers, metrology,
microlithography, medical imaging and surgery, wireless and optical
communications~\cite{he95,durnin87,salo99,durnin87a,davis00}.
Moreover, beams carrying definite photon OAM present novel internal
degrees of freedom that are potentially useful for quantum
information applications~\cite{molinaterriza02,molinaterriza07}.
Motivated by these issues, there has been an increasing interest in
generating and studying laser beams corresponding to particular
solutions of the scalar Helmholtz paraxial wave equation other than
the well known Hermite-Gaussian (HG) and the Laguerre-Gaussian (LG)
modes. Miller has solved the 3D Helmholtz paraxial wave equation in
17 coordinate systems, 11 of which were based on orthogonal
coordinates~\cite{miller}. Other examples of recently investigated
paraxial beams are Parabolic, Mathieu, Ince-Gaussian,
Helmholtz-Gaussian, Laplace-Gaussian, and pure light vortices~(see
\cite{kotlyar07} and references therein).

 In this Letter we introduce a novel family of paraxial beams which are solutions of
the scalar Helmoltz paraxial wave equation and are also eigenstates
of the photon OAM. The field profile of these beams is proportional
to the confluent hypergeometric function so we call them
Hypergeometric-Gaussian modes (HyGG). Unlike the hypergeometric
modes studied in~\cite{kotlyar07}, our HyGG modes carry a finite
power so that they can be realized in practical experiments.\\ The
field of the HyGG modes is given by
\begin{eqnarray}\label{eq:u}
    |HyGG\rangle_{pm} &=& u_{pm}(\rho,\phi;\zeta)=
      C_{pm} \frac{\Gamma\left(1+|m|+\frac{p}{2}\right)}{\Gamma\left(|m|+1\right)}\nonumber\\
   &&{\ensuremath\times}\, i^{|m|+1}\zeta^{\frac{p}{2}}(\zeta+i)^{-(1+|m|+\frac{p}{2})}\nonumber\\
   &&{\ensuremath\times}\,\rho^{|m|}e^{-\frac{i\rho^2}{(\zeta+i)}}e^{im\phi} \nonumber\\
   &&{\ensuremath\times}\,{}_{1}\!F_{1}\left(-\frac{p}{2}, |m|+1;\frac{\rho^2}{\zeta(\zeta+i)}\right)
\end{eqnarray}
where $m$ is integer, $p\ge-|m|$ is real valued, $\Gamma(x)$ is the
gamma function and ${}_{1}F_{1}(a,b;x)$ is a confluent
hypergeometric function~\cite{abramowitz}. In Eq.~(\ref{eq:u}) we
used dimensionless cylindrical coordinates $\rho=r/w_0$, $\phi$,
$\zeta=z/z_{R}$, where $w_0$ is the beam waist and $z_{R}=\pi
w_0^2/\lambda$ is the beam Rayleigh range. We notice the
characteristic factor $e^{im\phi}$ in the field, so that the integer
values of $m$ are identified with the eigenvalues of the photon OAM
in units of $\hbar$. The normalization
condition~\cite{inner_product} ${}_{pm}\langle
HyGG|HyGG\rangle_{pm}=1$ fixes the constant $C_{pm}$ to
$C_{pm}=[2^{p+|m|+1}/\pi\Gamma(p+|m|+1)]^{1/2}$ up to a constant
phase factor. The HyGG modes are an overcomplete not orthogonal set
of modes. The inner product of two normalized HyGG modes is given by
${}_{p'm'}\langle HyGG|HyGG\rangle_{pm}=\delta_{mm'}
\{\Gamma(p/2+p'/2+|m|+1)/[\Gamma(p'+|m|+1)\Gamma(p+|m|+1)]\}^{1/2}$.
The asymptotic behavior of the intensity
$|u_{pm}(\rho,\phi,\zeta)|^2$ of the HyGG modes as
$\rho\rightarrow\infty$ at fixed $\zeta>0$ is
$|u_{pm}|^2\propto\rho^{-2(2+p+|m|)}$ in general, and changes into
the Gaussian behavior
$|u_{pm}|^2\propto\rho^{2(p+|m|)}\exp[-2\rho^2/(1+\zeta^2)]$ when
$p$ is a non negative even integer (see below). At the beam center
($\rho\rightarrow 0$) the field of the HyGG modes vanishes as
$\rho^{|m|}$ as expected for the eigenmodes of the photon OAM (for
$\zeta \geq 0$). Because all zeroes of the hypergeometric function
${}_{1}F_{1}(a,b;x)$ occur for real values of $x$, a characteristic
feature of the HyGG modes given by Eq.~(\ref{eq:u}) is that their
intensity $|u_{pm}|^2$ never vanishes in the transverse plane,
except at the beam axis $\rho=0$ and at infinite. This confers to
the intensity of the HyGG modes the typical doughnut shape (a single
brilliant ring) for any values of $m$ and $p$, except $p=m=0$ when
the HyGG mode reduces to the TEM$_{00}$ Gaussian mode.\\ The limit
of the field $u_{pm}(\rho,\phi;\zeta)$ at the pupil plane
$\zeta\rightarrow 0^+$ is given by
\begin{equation}\label{eq:uz0}
  \lim_{\zeta\rightarrow0} u_{pm}(\rho,\phi;\zeta) =
                   C_{pm}\:\rho^{p+|m|}e^{-\rho^2+im\phi}.
\end{equation}
This feature is very useful because this pupil function can be
generated by applying the singular phase factor $e^{im\phi}$ to a
Gaussian-parabolic transmittance profile of the order $p+|m|$. In
particular, the HyGG modes with $p=-|m|$ are simply generated by
applying the phase factor $e^{im\phi}$ to a TEM$_{00}$ laser field at
its waist plane.\\ The HyGG modes can be expanded in the complete
basis of the Laguerre-Gauss (LG) modes. In general, the mode
HyGG$_{pm}$ is a superposition of the infinite LG$_{qm}$ modes with
same $m$ and any integer $q\geq 0$. In fact, when both the HyGG and
the LG modes are normalized, we have $|HyGG\rangle_{pm}
=\sum_{q=0}^\infty A_{pq} |LG\rangle_{qm}$ with coefficients $A_{pq}$
given by
\begin{equation}\label{eq:Apq}
   A_{pq} =\sqrt{\frac{(q+|m|)!}{q!\, \Gamma(p+|m|+1)}} \, \frac{\Gamma(q-p/2)\Gamma(p/2+|m|+1)}
                 {\Gamma(-p/2)\Gamma(q+|m|+1)}.
\end{equation}
The HyGG modes can exhibit different features when the mode parameters
$p$ and $m$ are changed. It is then convenient to separate the HyGG
modes in a few subfamilies having similar properties.
\begin{figure}[h]
     \begin{center}
        \includegraphics[width=8cm]{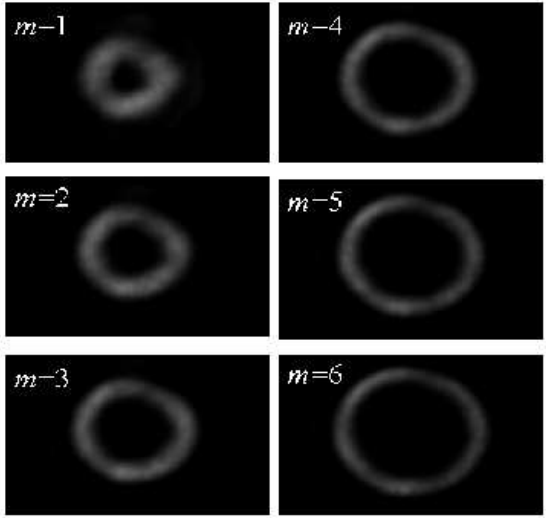}
     \end{center}
\caption{\label{fig:Fig1}Experimentally observed intensity
distributions of the $|HyGG\rangle_{-m,m}$ mode for $m=1,...,6$ in
the transverse plane $z=0.18 z_0$. The Rayleigh range was
$z_0=747.4$~cm.}
\end{figure}
\begin{enumerate}
\item
$p=m=0$\\ This mode is the well known Gaussian mode TEM$_{00}$.
\item
$p=-|m|$, odd.\\ The pupil field of these modes at $\zeta=0$ is
$\exp(-\rho^2+im\phi)$. At planes $\zeta>0$ the modes are linear
combinations of the modified Bessel functions $I_0(x)$ and $I_1(x)$,
where $x=\rho^2/2\zeta(\zeta+i)$. We may call this subfamily of modes
the modified Bessel Gauss (MBG) modes. Unlike the well known Bessel
modes, these modes carry a finite power and are not diffraction free.
When $\rho\rightarrow\infty$ at fixed $\zeta>0$, the intensity of
these modes vanishes according to $|u_{pm}|^2\propto\rho^{-4}$.
\item
$p=-|m|$, even.\\ The pupil field of these modes at $\zeta=0$ is
$\exp(-\rho^2+im\phi)$. At planes $\zeta>0$ the modes are linear
combinations of exponential $\rho$-dependent terms. We may call this
subfamily of modes the Modified Exponential Gauss (MEG) modes. When
$\rho\rightarrow\infty$ at fixed $\zeta$, the intensity of these modes
vanishes according to $|u_{pm}|^2\propto\rho^{-4}$.
\item
$p\ge0$, even.\\ The pupil field of these modes is given by
Eq.~(\ref{eq:uz0}). When $p$ is a non negative even integer, the
confluent hypergeometric function reduces to a Laguerre polynomial. We
will refer to these modes as to the modified Laguerre-Gauss modes
(MLG). The asymptotic behaviour of the intensity of the MLG modes as
$\rho\rightarrow\infty$ at fixed $\zeta>0$ is the same as for the
usual LG modes (i.\ e.\ $|u_{pm}|^2\propto
\rho^{2(p+|m|)}e^{-2\rho^2/(1+\zeta^2)}$). Unlike the LG$_{pm}$ modes, however, the
MLG$_{pm}$ modes have a single-ring intensity profile for any admitted
value of $p$. The MLG$_{pm}$ modes can be expressed as the linear
superposition of a finite number of LG$_{qm}$ modes, namely, the
LG$_{qm}$ modes having the same $m$ and integer $0\le q\le p/2$. In
fact, when $p$ is a non negative even integer, Eq.~(\ref{eq:Apq})
reduces to
\begin{equation}\label{eq:ApqLG}
  A_{pq} =(-1)^q \frac{(p/2)!\,(p/2+|m|)!}{(p/2-q)!\sqrt{q!\,(p+|m|)!\,(q+|m|)!}}\,
\end{equation}
where $0\le q\le p/2, p \mathrm{\ even}$. The quantities $A_{pq}$ form
the entries of a non singular
$(p/2+1){\ensuremath{\ensuremath{\ensuremath\times}}}(p/2+1)$ matrix. It is then obvious
that this sub-family of HyGG modes forms a complete, yet not
orthogonal, set of functions in the transverse plane and that the full
set of HyGG modes is therefore overcomplete.
\end{enumerate}

In our experiment, a TEM$_{00}$ linearly polarized laser beam from a
frequency doubled \textit{Nd:YVO}$_4$ ($\lambda=532$~nm, Model Verdi
V5, Coherent) was used to illuminate a grey scale computer generated
hologram (CGH) sent onto the LCD microdisplay of a spatial light
modulator (SLM) (HoloEye Photonics LC-R 3000), with
$1920{\ensuremath\times}1200$ pixels in a rectangle
18.24{\ensuremath\times}11.40~mm wide. The SLM was located in the
waist of the incident beam. We performed two series of measurements,
according to the beam waist values $w_0$ at the SLM position. We
measured $w_0$ by best fit between the error function and the
integral curve of the intensity profile of incident beam as seen by
CCD camera.The measured values were
$w_0=1.13{\ensuremath\pm}0.04$~mm and
$w_0=0.115{\ensuremath\pm}0.004$~mm, corresponding to Rayleigh
ranges $z_0=747.4{\ensuremath\pm}50$ cm and
$z_0=7.8{\ensuremath\pm}0.5$~cm, respectively. The theory was
compared with the experimental data just using as parameters the
measured values reported above, without further best fitting. We
focused our attention on the HyGG modes with $p=-|m|$.
\begin{figure}[htb]
    \begin{center}
        \includegraphics[width=6cm]{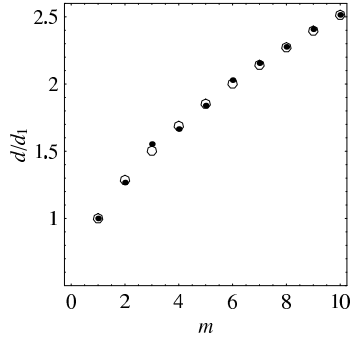}
     \end{center}
\caption{\label{fig:Fig2} The ring diameter $d$ of the
$|HyGG\rangle_{-m,m}$ mode for $m=1,...,10$ measured at plane
$z=0.18 z_0$. The Rayleigh range was $z_0=747.4$~cm. The reported
values of the diameters were scaled with respect to the value $d_1$
for $m=1$. \protect$\circ$ -- theory, \protect$\bullet$ --
experiment.}
\end{figure}
In accordance with the theoretical predictions, we observed an
intensity profile in the transverse plane essentially made of a
single bright annulus, whatever the value of $m$ we used or the
observation $z$-plane. Some instances of the observed intensity
profiles are shown in Fig.~\ref{fig:Fig1}. The ring diameter of the
beam as a function of $m$ is reported in Fig~\ref{fig:Fig2}. The
diameter $d$ of the ring was defined as the maximum distance between
any two opposite maxima of the intensity profile. The scaling law of
$d$ versus $m$ turned to be in good agreement with the theoretical
predictions. We measured also the ratio between the diameter $d(z)$
of the luminous ring and the gaussian beam size $w(z)$ at different
$z$ planes. The measurements were made by switching on and off the
CGH to compare the intensity profile of the HyGG mode with the
gaussian profile TEM$_{00}$ beam profile at the same plane.
Figure~\ref{fig:Fig3} shows that the ratio $d(z)/w(z)$ was the same
in all $z$ planes, as predicted by theory when $z>z_0$. The constant
value of the ratio $d/w$ obtained from the experiment was
$3.1{\ensuremath\pm}0.3$, which is close to the theoretical
prediction
$d/w = 3.6$.\\
\begin{figure}[htb]
    \begin{center}
         \includegraphics[width=6cm]{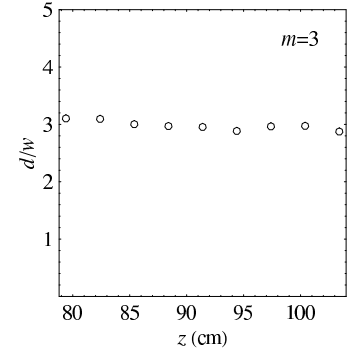}
    \end{center}
\caption{\label{fig:Fig3} The ratio between the diameter $d(z)$ of
the $|HyGG\rangle_{-3,3}$ mode and the $1/e^2$ intensity radius $w(z)$
of the generating TEM$_{00}$ gaussian beam as a function of $z$. The
Rayleigh range was $z_0 = 7.8$~cm.}
\end{figure}
In summary, we studied a novel family of paraxial beams having
hypergeometric field profile. This set of modes is overcomplete and
nonorthogonal and all modes carry a finite power. In spite of their
complicated field profile, these mode have a simple profile at the
pupil plane. Finally we have experimentally produced these modes for
different values of indexes. The agreement between experimental
results and theoretical predictions turned to be satisfactory.

%

\begin{thebibliography}{4}

\bibitem{he95}
H.~He, N.~R. Heckenberg, and H.~Rubinsztein-Dunlop, J.\ Mod.\ Opt. \textbf{42},
  217 (1995).

\bibitem{durnin87}
J.~Durnin, J.~J.~M. Jr., and J.~H. Eberly, Phys.\ Rev.\ Lett. \textbf{58}, 1499
  (1987).

\bibitem{salo99}
J.~Salo, J.~Fagerholm, A.~T. Friberg, and M.~M. Salomaa, Phys.\ Rev.\ Lett.
  \textbf{83}, 1171 (1999).

\bibitem{durnin87a}
J.~Durnin, J.\ Opt.\ Soc.\ Am.~A \textbf{4}, 651 (1987).

\bibitem{davis00}
J.~A. Davis, D.~E. McNamara, D.~M. Cottrell, and J.~Campos, Opt.\ Lett.
  \textbf{25}, 99 (2000).

\bibitem{molinaterriza02}
G.~Molina-Terriza, J.~P. Torres, and L.~Torner, Phys.\ Rev.\ Lett. \textbf{88},
  013601 (2002).

\bibitem{molinaterriza07}
G.~Molina-Terriza, J.~P. Torres, and L.~Torner, Nat.\ Phys. \textbf{3}, 305
  (2007).

\bibitem{miller}
W.~{Miller Jr.}, \emph{Symmetry and Separation of Variables} (Addison-Wesley,
  Boston, MA, 1977).

\bibitem{kotlyar07}
V.~V. Kotlyar, R.~V. Skidanov, S.~N. Khonina, and V.~A. Soifer,
Opt.\ Lett.
  \textbf{32}, 742 (2007).

\bibitem{abramowitz}
M.~Abramowitz and I.~A. Stegun, \emph{Handbook of Mathematical Functions}
  (Dover Publications Inc., New York, 1970).

\bibitem{inner_product}
The Hilbert inner product is defined according to $\langle
u|v\rangle = \int\!\! u^* v\rho d\rho d\phi$.

\end{thebibliography}
\end{document}